\documentclass[review]{elsarticle}
\usepackage{color}
\usepackage{graphicx}
\usepackage{lineno,hyperref}
\modulolinenumbers[5]
\journal{Journal of Crystal Growth}
\begin{document}

\begin{frontmatter}

\title{Utilization of mechanical alloying method for flux growth of single crystalline BaFe$_2$(As$_{1-x}$P$_x$)$_2$}
\author{Kosuke Z. Takahashi}
\author{Daisuke Okuyama}
\author{Taku J. Sato\corref{mycorrespondingauthor}}
\cortext[mycorrespondingauthor]{Corresponding author}
\ead{taku@tagen.tohoku.ac.jp}

\address{Institute of Multidisciplinary Research for Advanced Materials, Tohoku University, 2-1-1 Katahira, Sendai 980-8577, Japan}

\begin{abstract}
Mechanical alloying method has been employed to prepare the Ba-Fe-As-P precursors, necessary for the Ba-(As,P) flux growth of the single crystalline BaFe$_2$(As$_{1-x}$P$_x$)$_2$.
By alloying constituent elementals mechanically, the Ba-(As,P) precursors are successfully formed at the room temperature within one hour, significantly reducing preparation time.
Using the mechanically alloyed precursors, we have grown single crystals of BaFe$_2$(As$_{1-x}$P$_x$)$_2$ with the sizes up to 5~mm $\times$ 5~mm $\times$ 0.1~mm.
\end{abstract}


\end{frontmatter}

%\ead{taku@tagen.tohoku.ac.jp}

%\begin{keyword}
%Iron superconductor, crystal growth, mechanical alloying
%\end{keyword}

%\end{frontmatter}

%\linenumbers

\section{Introduction}
Discovered in 2008~\cite{doi:10.1021/ja800073m,1468-6996-16-3-033503}, iron-based superconductors have attracted a tremendous interest because of their high superconducting transition temperatures $T_{\rm c}$ up to 55~K~\cite{doi:10.1209/0295-5075-83-17002,doi:10.1143/JPSJ.77.063707,doi:10.1103/PRB.84.024521}.
Phosphorus-doped BaFe$_2$(As$_{1-x}$P$_x$)$_2$ is one of such iron-based superconductors with the maximum  $T_{\rm c}~{\rm (onset)} \leq 31$~K~\cite{Jiang09, Nakai10, PhysRevB.81.184519}.
The phosphorus ion is isovalent to the arsenic ion, and hence, the P-doping will not change charge-carrier concentrations. Instead, the P-doping is believed to introduce chemical pressure.
In addition, the disorder induced by the P-doping localizes to the (As,P)-site (chemical disorder), and hence less impurity scattering may be expected in the two-dimensional square lattice of Fe atoms.
Consequently, the Fe square lattice is believed to be much cleaner than those in the widely studied electron-doped compound Ba(Fe,Co)$_2$As$_2$~\cite{PhysRevLett.101.117004,PhysRevLett.101.207004}.
Recently, an intriguing electronic nematic phase was suggested above its structural phase transition at $T_{\rm s}$~\cite{Kasahara12}, further enhancing research activities on this P-doped iron superconductor~\cite{0953-2048-25-8-084005}.

A large-sized and high-quality single crystal is crucial for reliable physical property measurements.
To date, such a single crystal of BaFe$_2$(As$_{1-x}$P$_x$)$_2$ is mainly obtained using the flux growth method~\cite{PhysRevLett.105.207004,PhysRevB.85.184525,Nakajima12,Chen14}.
For this method, the precursors, Ba$_{2}$As$_{3}$, Ba$_{2}$P$_{3}$, FeAs and FeP, are first prepared by the low-temperature solid-state reaction, in order to avoid explosions due to the high vapor pressure of the elemental P.
Only after obtaining precursors, the high-temperature flux growth may be performed.
The low-temperature solid-state reaction and the two-step heat treatment are time-consuming, and hence an efficient method to prepare the precursors has been highly desired to reduce total duration of the sample preparation.

Recently, Weiss {\it et al.} reported the preparation of polycrystalline samples of potassium doped (Ba,K)Fe$_2$As$_2$ by the mechanical alloying method~\cite{weiss13}.
The mechanical alloying method is advantageous for preparing high-vapor-pressure materials, since the reaction takes place at the room temperature~\cite{Suryanarayana01}.
In this study, we have employed the mechanical alloying method for the preparation of the phosphorus doped BaFe$_2$(As$_{1-x}$P$_x$)$_2$.
The precursors for the flux growth [{\it i.e.}, BaFe$_2$(As$_{1-x}$P$_x$)$_2$ + Ba-(As,P) flux] were successfully obtained at the room temperature in a single mechanical alloying run within an hour.
The precursors were used for the subsequent flux growth, and single crystals were successfully grown with the sizes up to 5~mm $\times$ 5~mm $\times$ 0.1~mm.

\section{Experimental details}
Elemental Ba (99.9~\%), Fe (99.9~\%), As (99.9999~\%) and P (99.9999~\%) were used as starting materials.
For the preparation of the precursors, starting elemental materials with the molar ratio~\cite{Nakajima12} Ba:Fe:As:P = 6:1:10(1-$x_{\rm nom}$):10$x_{\rm nom}$ were loaded in the stainless steel pot (30 mm diameter, 35 mm height) with eight 10~mm diameter stainless steel balls.
$x_{\rm nom}$ stands for the nominal ratio of P to As.
Room temperature mechanical alloying was performed for an hour at the speed of 400~rpm using the planetary ball mill (Fritsch P-7).
Resulting precursors were then loaded in an Al$_2$O$_3$ crucible, sealed in a quartz tube with inert Ar gas of the pressure being 0.25~atm at the room temperature.
The tube was heated up to 1423~K for 24~h, kept at the temperature for 3~h, and then slowly cooled down to 1173~K with the cooling rate of 1~K/h.
The grown single crystals were mechanically removed from the flux.
All the sample preparation procedures were performed in the glove box filled with the inert Ar gas with the oxygen concentration of about 1~ppm.

The microstructure and composition of the obtained single crystals were checked by the scanning electron microscope (SEM; Hitachi-S-4800) equipped with the energy dispersive x-ray (EDX) analyzer.
The quality of the obtained crystals was checked by the four-circle x-ray diffractometer (MAC Science M18X) with the Mo K$\alpha$ radiation.
The magnetization of the resulting single crystals was measured using the commercial superconducting-quantum-interference-device magnetometer MPMS-XL (Quantum Design).
Electrical resistivity was measured using the standard four-probe method with the excitation current $I = 1$~mA.

\section{Results and discussion}

\begin{figure}
 \includegraphics[scale=0.3, angle=0, trim={0cm 3cm 0cm 0cm}]{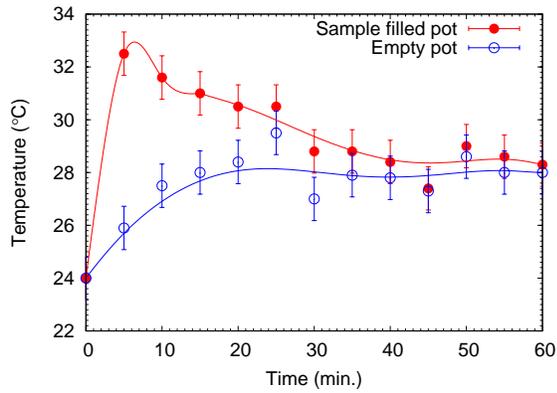}
 \vspace*{0.5cm}
  \caption{Evolution of the pot temperatures during the mechanical alloying process for $x_{\rm nom} = 0.45$.
    Red-filled circles stand for the temperature of the pot containing the starting materials, whereas blue-open  circles are for the empty pot.
    Solid lines are guides to the eyes.
  }
\label{fig:tempChange}
\end{figure}

\begin{figure}
  \includegraphics[scale=0.20, angle=0, trim={0cm 7cm 12cm 15cm}]{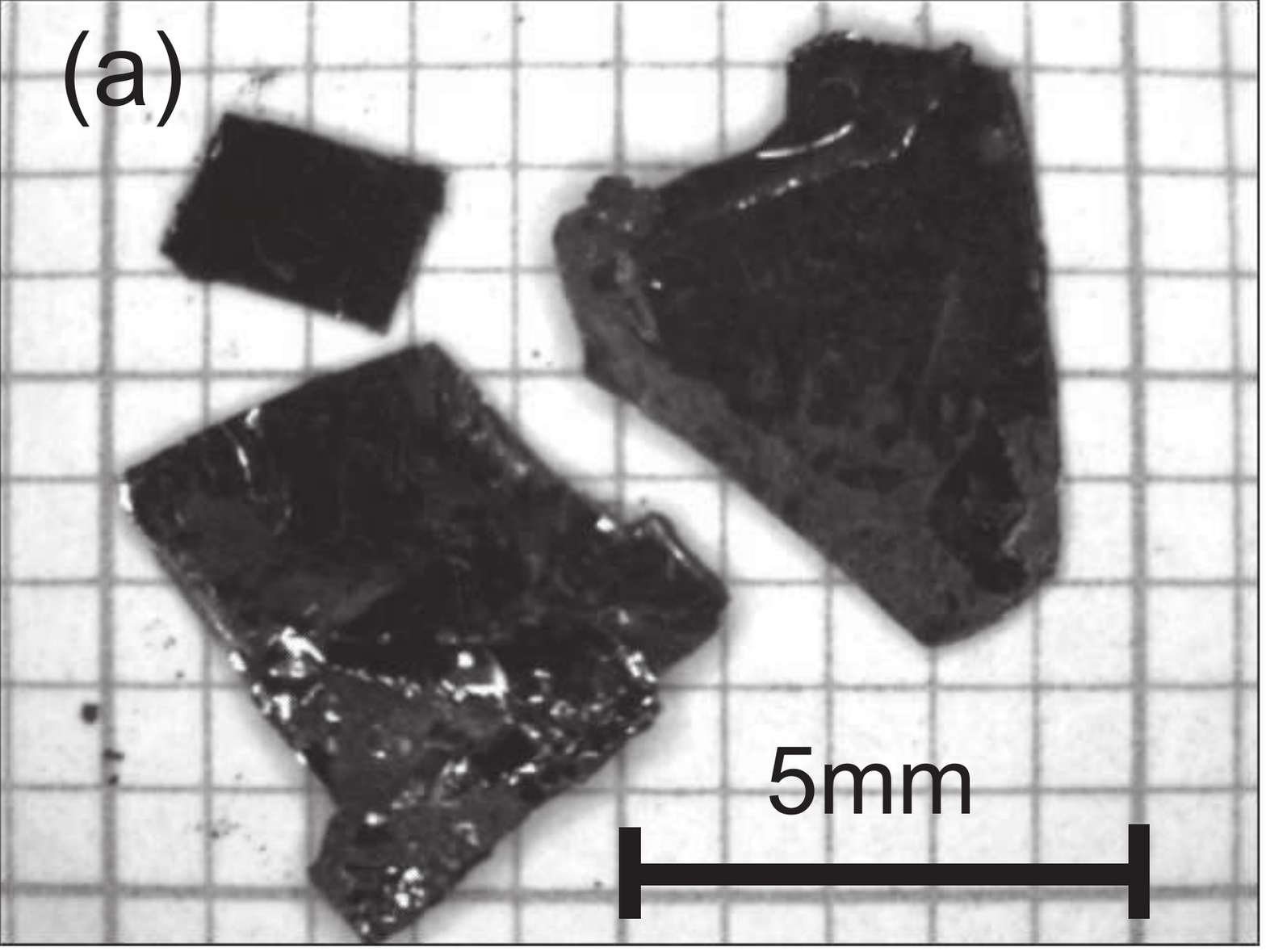}\
  \includegraphics[scale=0.35, angle=0, trim={0cm 2cm 0cm 15cm}]{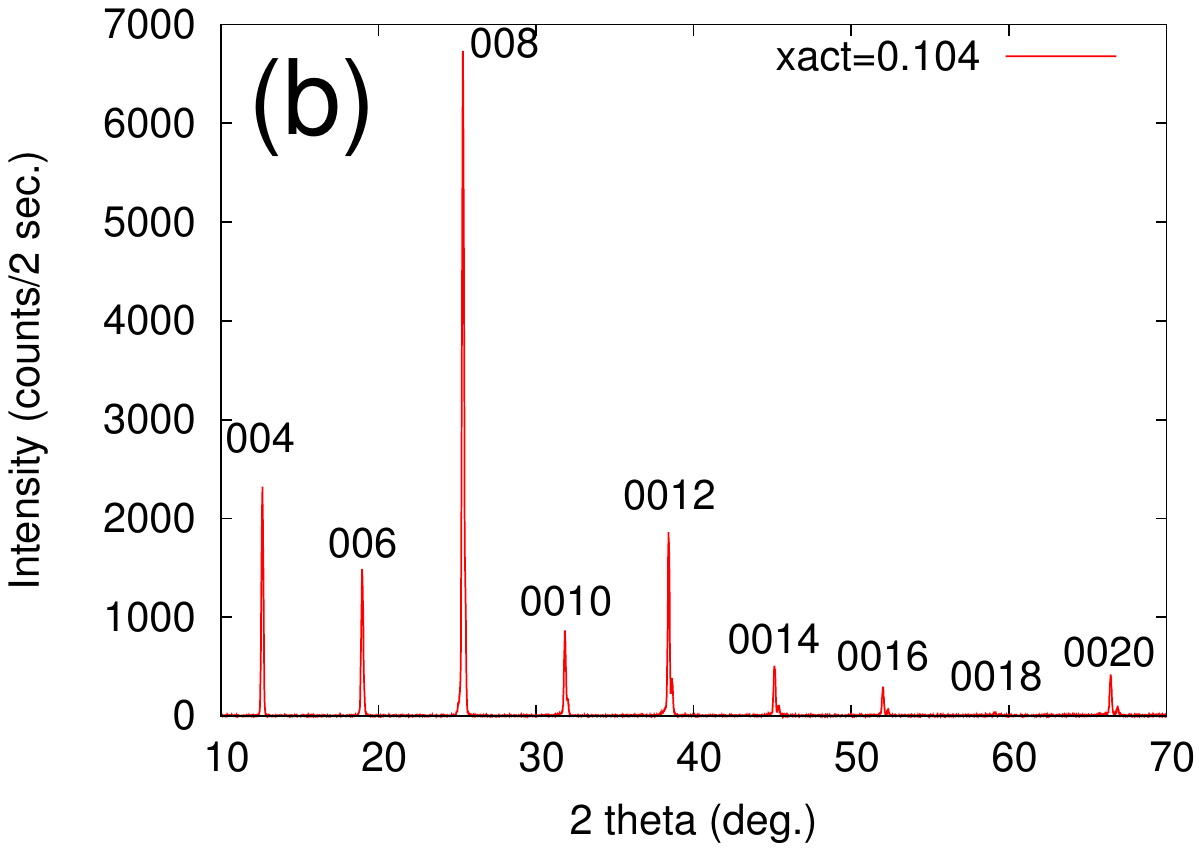}
  \caption{(a) Optical micrograph of the obtained BaFe$_2$(As$_{0.90}$P$_{\rm 0.10}$)$_2$ ($x_{\rm act} = 0.104$; $x_{\rm nom} = 0.015$) single crystals.
    (b) Single crystal x-ray diffraction pattern along the $00l$ direction for $x_{\rm act} = 0.104$.
  }
\label{fig:singleQuality}
\end{figure}

First, elemental mixtures with $x_{\rm nom} = 0.015, 0.040, 0.050, 0.100, 0.150, 0.200, 0.330, 0.450$, and $0.500$ were mechanically alloyed at the room temperature.
Typical behavior of the pot temperatures during the mechanical alloying process for $x_{\rm nom} = 0.45$ is shown in Fig.~1.
The temperature difference between the filled and empty pots can be regarded as the temperature increase due to the exothermic reaction.
The temperature difference increases rapidly within the first 5 minutes, indicating that the reaction mostly completes in this period.
For complete reaction, we performed the mechanical alloying for an hour for all the compositions. 
The mechanical alloying yielded black-colored fine powders.
X-ray powder diffraction of the obtained powder samples (not shown) detected reflections from the BaFe$_2$(As,P)$_2$ phase, in addition to several unidentified peaks.
We could not see peaks from elemental P, As, Fe and Ba, and hence we can conclude that some pseudo-binary Ba-(As,P) or pseudo-ternary Ba-Fe-(As,P) precursors are formed, in addition to the target BaFe$_2$(As,P)$_2$ compound.

Next, we performed the flux growth using thus-obtained precursors.
As a result, we have successfully grown thin-plate-shaped single crystals up to 5~mm $\times$ 5~mm $\times$ 0.1~mm.
It may be noted that the temperature was increased directly to the highest temperature 1423~K, without having a low-temperature solid-state reaction.
An optical micrograph of the grown crystals for $x_{\rm nom}$ = 0.015 is shown in Fig.~\ref{fig:singleQuality}(a).
Shiny facet can be seen in the micrograph, corresponding to the tetragonal $ab$ plane of the TrCr$_2$As$_2$ structure (space group $I4/mmm$)~\cite{Pfisterer80}.

The compositions of grown crystals were checked by the EDX analysis, and are summarized in Table~\ref{tb:compositionEDX}.
The actual composition $x_{\rm act}$ rapidly increases with increasing the nominal composition up to $x_{\rm nom} \simeq 0.2$.
For $x_{\rm nom} \geq 0.2$, $x_{\rm act}$ shows weaker increase with ${\rm d}x_{\rm nom}/{\rm d}x_{\rm act} \simeq 1$.
This is consistent with the previous report~\cite{Nakajima12}, and suggests different crystal growth mechanisms for the two ranges of the composition, $x_{\rm nom} < 0.2$ and $x_{\rm nom} > 0.2$.

X-ray diffraction was measured along the 00$l$ direction for the obtained crystals using the reflection geometry.
A representative result for $x_{\rm act} = 0.104$ is shown in Fig.~\ref{fig:singleQuality}(b).
Sharp peaks were observed, obeying the reflection condition of $I4/mmm$.
This confirms single crystal nature of the obtained crystal.
The lattice parameter $c$ was estimated for selected $x_{\rm act}$ by fitting the diffraction profile in a range $4\leq l \leq 20$.
The resulting $c$ is given in Table~\ref{tb:compositionEDX}.
The linear relation between $c$ and $x_{\rm act}$ was found, and was approximated as $c = -0.57(3)x_{\rm act} + 13.02(2)$.
This confirms the Vegard's law in this solid solution, being consistent with previous reports~\cite{Jiang09,PhysRevB.81.184519}.

\begin{figure}[t]
  \centering
  \includegraphics[scale=0.3, angle=0]{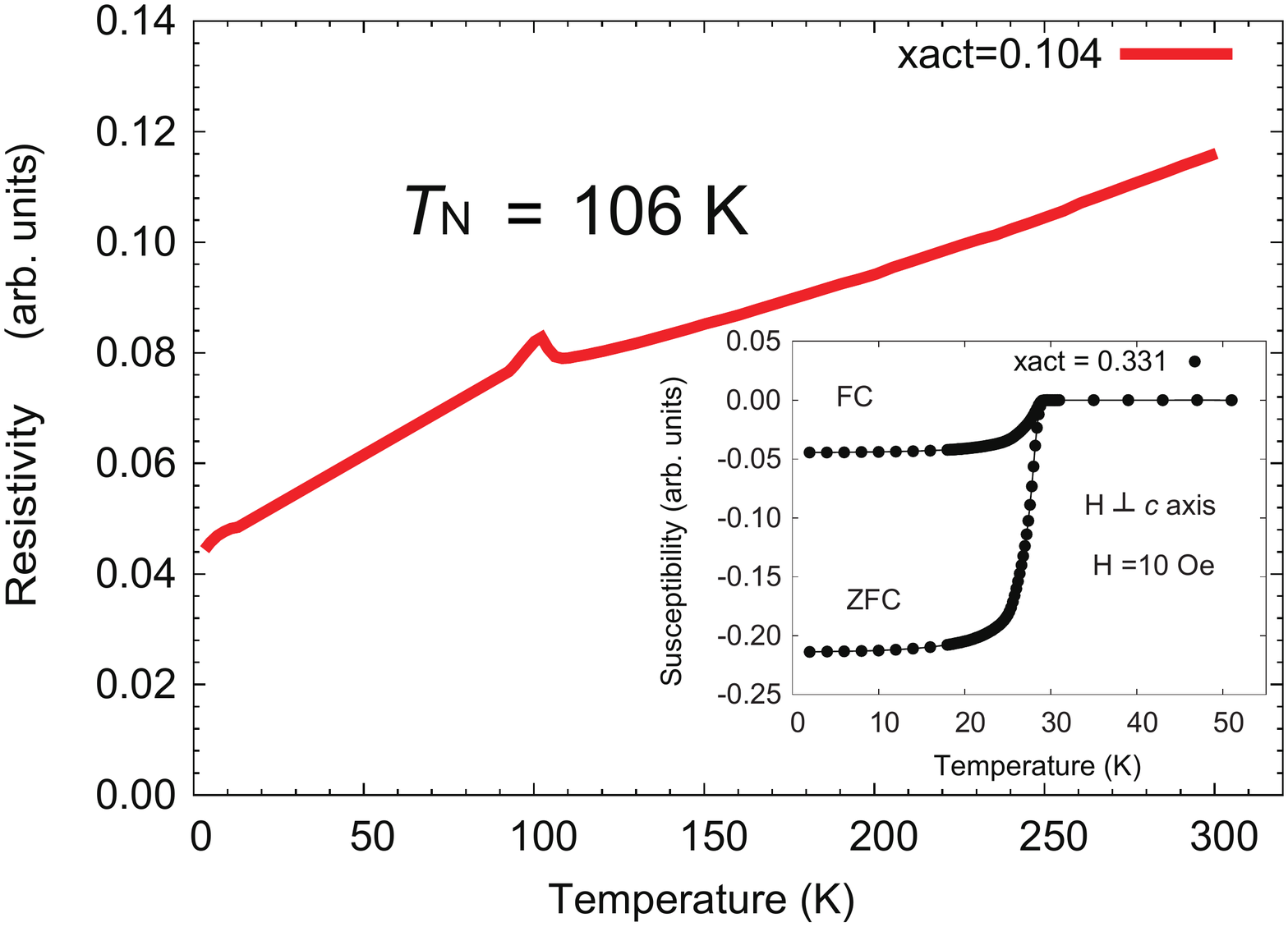}
  \caption{Temperature dependence of the resistivity $\rho$ for $x_{\rm act} = 0.104$.
    The anomaly observed at $T_{\rm N} = 106$~K corresponds to the antiferromagnetic transition.
    Inset: temperature dependence of the magnetic susceptibility for the superconducting $x_{\rm act} = 0.331$ crystal.
    Field-cooled (FC) and zero-field-cooled (ZFC) results are shown.
    External magnetic field of 10~Oe was applied perpendicular to the $c$ axis.
  }
\label{fig:singleSusRes}
\end{figure}

The electrical resistivity was measured for selected crystals.
As a typical example, the result for $x_{\rm act} = 0.104$ is shown in Fig.~\ref{fig:singleSusRes}.
A clear anomaly in the resistivity was observed at $T_{\rm N} = 106$~K.
This corresponds to the antiferromagnetic transition, and $T_{\rm N}$ is in a good agreement with the earlier reports~\cite{Jiang09,Nakajima12}.
The residual resistivity ratio (RRR) is approximately 2.6, estimated from the resistivity at 300~K and 3.25~K.
This RRR value is comparable to the value for the as-grown $x_{\rm act} = 0.13$ crystal in the previous report (RRR $\simeq 2.4$)~\cite{Nakajima12}.
Magnetic susceptibility was measured down to 2~K under the external magnetic field  $H = 10$~Oe applied perpendicular the $c$ axis.
The result for the $x_{\rm act} = 0.331$ crystal is shown in the inset of Fig.~\ref{fig:singleSusRes}.
The superconducting transition was clearly observed at $T_{\rm c} = 27.6$~K (midpoint).
The transition temperature is again in a good agreement with the earlier reports~\cite{Jiang09,Nakajima12}.
The sharpness of the susceptibility drop confirms the compositional homogeneity of the grown crystal.
From these results, we conclude that mostly the same quality single crystal can be grown using the precursors prepared by the mechanical alloying method.

\begin{table}
\begin{center}
\begin{tabular}{ccccccccc}
$x_{\rm nom}$ & Ba~(\%) & Fe~(\%) & As~(\%) & P~(\%) & $x_{\rm act}$ & $c$ (\AA) & $T_{\rm N}$ (K) & $T_{\rm c}$ (K) \\
\hline \hline
0.015 & 21.2(1) & 38.6(2) & 36.1(1) & 4.2(1) & 0.104(3) & 12.9551(2) & $ 106 $ & - \\
0.040 & 20.9(1) & 38.9(2) & 27.6(1) & 12.6(1) & 0.313(2) & 12.8291(1) & - & $ 27.3 $ \\
0.050 & 22.2(2) & 40.4(4) & 25.1(2) & 12.4(2) & 0.331(3) & 12.8148(8) & - & $ 27.6 $ \\
0.100 & 20.7(2) & 37.5(2) & 19.7(6) & 22.2(7) & 0.53(1) & 12.7578(2) & - & $ 13.9 $ \\
0.150 & 20.6(1) & 37.7(1) & 12.4(1) & 29.4(1) & 0.704(2) & 12.6145(4) & - & - \\
0.200 & 21.11(7) & 38.1(2) & 9.5(1) & 31.3(2) & 0.768(2) & - & - & - \\
0.330 & 20.0(2) & 35.8(3) & 8.0(2) & 36.3(2) & 0.820(4) & - & - & -  \\
0.450 & 20.7(2) & 37.6(3) & 3.7(1) & 37.9(4) & 0.910(3) & 12.4818(4) & - & -  \\
0.500 & 20.6(1) & 36.72(7) & 1.63(8) & 41.1(2) & 0.962(2) & - & - & - \\
\hline
\end{tabular}
\end{center}
\caption{Compositions of the grown crystals determined by the EDX analysis. 
  Each elemental composition is the averaged value of five independent measurements of the same crystal surface, and its standard deviation is given in the parentheses.
The lattice constant $c$ is estimated from the x-ray diffraction experiment.
$T_{\rm c}$ is determined from the magneic susceptibility drop (midpoint), whereas $T_{\rm N}$ from the temperature of the inflection point in the resistivity.
}
\label{tb:compositionEDX}
\end{table}

\section{Summary}
In summary, we have used the mechanical alloying method to prepare the precursors for the flux growth of BaFe$_2$(As$_{1-x}$P$_x$)$_2$ single crystals.
The precursors were successfully obtained at the room temperature within an hour.
Using the mechanically alloyed precursors, the single crystals were grown with the sizes up to 5~mm $\times$ 5~mm $\times$ 0.1~mm in the P concentration range $0 < x_{\rm act} < 1$.
The x-ray diffraction, magnetic susceptibility and electrical resistivity measurements show that the obtained crystals are of the same quality as the crystals obtained in the standard manner~\cite{Nakajima12}.

\section{Acknowledgments}
The authors thank A.~P.~Tsai and S.~Kameoka for allowing us to use their planetary ball mill, T. Masuda and T. Haku for allowing us to use their high-energy x-ray instrument, and S. Ohashi and M. Ishii for technical support in SEM measurements.
This work is partly supported by JSPS Grants-in-Aid for Scientific Research (B) 23340097 and (A) 23244068.

%\section*{References}
%%\bibliographystyle{prsty}
\bibliography{bafe2asp2}

\end{document}